# Dynamic Mode Decomposition for data-driven analysis and reduced-order modelling of E×B plasmas: II. dynamics forecasting

F. Faraji*[1], M. Reza*, A. Knoll*, J. N. Kutz**

* Plasma Propulsion Laboratory, Department of Aeronautics, Imperial College London, London, United Kingdom

** Department of Applied Mathematics and Electrical and Computer Engineering, University of Washington, Seattle, United States

**Abstract**: In part I of the article, we demonstrated that a variant of the Dynamic Mode Decomposition (DMD) algorithm based on variable projection optimization, called Optimized DMD (OPT-DMD), enables a robust identification of the dominant spatiotemporally coherent modes underlying the data across various test cases representing different physical parameters in an E×B simulation configuration. We emphasized that the OPT-DMD significantly improves the analysis of complex plasma processes, revealing information that cannot be derived using conventionally employed analyses such as the Fast Fourier Transform. As the OPT-DMD can be constrained to produce stable reduced-order models (ROMs) by construction, in this paper, we extend the application of the OPT-DMD and investigate the capabilities of the linear ROM from this algorithm toward forecasting in time of the plasma dynamics in configurations representative of the radial-azimuthal and axial-azimuthal cross-sections of a Hall thruster and over a range of simulation parameters in each test case. The predictive capacity of the OPT-DMD ROM is assessed primarily in terms of short-term dynamics forecast or, in other words, for large ratios of training-to-test data. However, the utility of the ROM for long-term dynamics forecasting is also presented for an example case in the radial-azimuthal configuration. The model's predictive performance is heterogeneous across various test cases. Nonetheless, a remarkable predictiveness is observed in the test cases that do not exhibit highly transient behaviors. Moreover, in all investigated cases, the error between the ground-truth and the reconstructed data from the OPT-DMD ROM remains bounded over time within both the training and the test window. As a result, despite its limitation in terms of generalized applicability to all plasma conditions, the OPT-DMD is proven as a reliable method to develop low computational cost and highly predictive data-driven reduced-order models in systems with a quasi-periodic global evolution of the plasma state.

**Section 1: Introduction**

Today, reliable, predictive, and generalizable reduced-order models do not exist for plasmas. There are at least two reasons for this status quo: first, the classic conservation equations derived from the moments of the plasma kinetic equation [1] do not include the important effects of microscopic plasma instabilities and oscillations on the electrons' momentum and energy transport [2]. Second, despite years of effort and several approaches pursued [3]-[7], rigorous and generalizable closure models for the conservation equations are still to be established so that the effects of the kinetic phenomena and processes such as the cross-field electrons' transport can be self-consistently resolved in reduced-order simulations based upon the conservation equations for the plasma.

Nonetheless, the need for self-consistent, interpretable reduced-order plasma models is critical for scientific and industrial advancements alike. From an academic perspective, the availability of such models can enable answering the so-far unresolved questions in the physics of cross-field plasmas, particularly with regards to the excitation and evolution of the plasma instabilities and turbulence as well as their interactions with plasma species that, for example, can result in enhanced transport of the particles and energy across the magnetic field. From an applied point of view, reliable reduced-order models can lead to the prediction and control of the plasmas, paving the way for more efficient technological solutions and novel plasma applications.

The above issues, although rather different in nature and extent, also exist in other research fields such as fluid mechanics. Attempts to establish closure models for Navier-Stokes system of equations to incorporate the effects of unresolved turbulence, for example, have been rigorously pursued for decades in order to achieve fully generalizable predictive models of the fluid systems [8]. Nonetheless, the efforts in fluid dynamics to the above end have not been fully successful either.

In recent years, however, with the rapid and remarkable advances in the computational capacity, on the one hand, and in the data-driven (DD)/machine-learning (ML) algorithms, on the other, new pathways have been proposed that aim to utilize the capabilities of the DD/ML techniques and the rapidly increasing body of data to finally enable generalizable and predictive reduced-order models in various applications, from fluid systems for which the underlying system of equations is partially known, to other dynamical systems for which no known governing equations exist [9]. These efforts have led to highly interesting outcomes and, despite generalizability and interpretability still persist as the main challenges to overcome [10], the potentials of several ML algorithms for engineering and physics modelling are well demonstrated [9].

---

[1] **Corresponding Author** (f.faraji20@imperial.ac.uk)

Two of the most promising application areas of the ML algorithms are: (1) dimensionality reduction of high-dimensional, high-fidelity data and the extraction of the underlying dominant patterns [9][10][11], and (2) the development of predictive reduced-order PDE/ODE models [12][13]. In part I of this article [14], we extensively examined in different plasma setups the potentials of the Dynamic Mode Decomposition (DMD) in the first application area. We investigated the utility of this method toward an augmented analysis of the high-fidelity simulations data and the explanation of the complex behaviors.

In the present part, we turn our focus to the second area. We assess the applicability of a variant of the DMD based on variable projection optimization [15] to derive reliable reduced-order models for plasma systems in terms of the time evolution of the dominant spatiotemporally coherent structures in the data. The target DMD variant, referred to as OPT-DMD [15], was introduced in part I [14]. It was, importantly, pointed out that the development of OPT-DMD [15] and bagging optimized DMD (BOP-DMD) [16] has significantly changed the potential of the DMD to perform critical modeling tasks due to the robustness of these new methods to noise. Moreover, OPT-DMD was emphasized to enable constraining best-fit linear models to be stable by construction (eigenvalues are constrained to the left-half plane), purely oscillatory (eigenvalues are constrained to the imaginary axis), or complex conjugate pairs (solutions are real). Thus, OPT-DMD provides a viable strategy that is robust and stable for real-world applications and datasets.

As a result, the emphasis here is primarily on gauging the extent of predictiveness of the linear models from the OPT-DMD algorithm across a set of sample plasma conditions with varying degrees of nonlinearity in the underlying processes and interactions. Since this article serves as the first attempt to assess the applicability of the OPT-DMD method to develop models from real data of complex plasma systems, it is deemed most appropriate to focus on the qualitative performance of the data-driven models. In this respect, devising possible strategies for automated fine-tuning of the hyper-parameters in the OPT-DMD algorithm, particularly the truncation rank ($r$) [14], will be the next step in building upon the results of this effort which is to be pursued in future work.

The high-fidelity ground-truth data used to derive and test the OPT-DMD models are obtained from the reduced-order quasi-two-dimensional PIC simulations of the test cases that will be overviewed in Section 2. Developed at Imperial Plasma Propulsion Laboratory (IPPL), the computationally efficient reduced-order PIC code [17]-[19] plays an enabling role in our research toward data-driven predictive plasma modelling. This is because of the proven capability of this PIC scheme to provide large sets of high-fidelity data over extended parameter spaces and time windows at a significantly reduced computational resource compared to traditional multi-dimensional PIC codes [17][20][21].

The remainder of this article is structured as follows: In Section 2, we provide an overview of the test cases, which correspond to E×B plasma configurations representative of the axial-azimuthal and radial-azimuthal cross-sections of a Hall thruster. The OPT-DMD is applied to the axial-azimuthal and radial-azimuthal simulations' data over a range of sample setup parameters. The predictive performance of the OPT-DMD linear model in each test case is presented and discussed in Section 3.

**Section 2: Overview of the test case high-fidelity simulations**

The setups of the test-case simulations in this work follow, in general, those of the well-established benchmarks available in the literature in the axial-azimuthal and radial-azimuthal Hall thruster coordinates [22][23]. In our previous publications, we verified the reduced-order PIC code against these benchmark cases [17][18][20] and used the radial-azimuthal benchmark setup to perform a parametric study on the influence of several physical factors on the physics of instabilities and electron transport [21]. We complemented the data from our previous studies with new data from recently carried out simulations in order to create an expanded set of high-fidelity data that span over a broader range of simulation conditions. An overview of the setup for the radial-azimuthal and axial-azimuthal test cases are presented in Sections 2.1 and 2.2, respectively.

## 2.1. Radial-azimuthal cross-field discharge configuration

The setup of the radial-azimuthal test cases was overviewed in part I of the article [14]. As pointed out in part I, the test-case simulations feature a temporally constant cosine-shaped ionization source along the radial direction with the peak value of $S$ to compensate for the flux of particles lost to the walls. This ionization source has been implemented as described in Refs. [20][23]. Furthermore, in all cases, a constant axial electric field ($E$) and a constant radial magnetic field ($B$) are also applied.

The various radial-azimuthal test cases in this part II differ from each other based on the set of values chosen for the ionization source peak, the axial electric field, and the radial magnetic field. Table 1 reports the values of these physical parameters for each test case that are presented and discussed in Section 3.

| Test case | S value | E value | B value |
|---|---|---|---|
| 1 (Baseline) | $S_0 = 8.9 \times 10^{22}\ m^{-3}s^{-1}$ | $E_0 = 10{,}000\ Vm^{-1}$ | $B_0 = 0.02\ T$ |
| 2 | $S_0$ | $0.5E_0$ | $B_0$ |
| 3 | $S_0$ | $3E_0$ | $B_0$ |
| 4 | $S_0$ | $5E_0$ | $B_0$ |
| 5 | $S_0$ | $E_0$ | $1.5B_0$ |
| 6 | $1/16\ S_0$ | $E_0$ | $B_0$ |
| 7 | $1/8\ S_0$ | $E_0$ | $B_0$ |
| 8 | $3S_0$ | $E_0$ | $B_0$ |
| 9 | $6S_0$ | $E_0$ | $B_0$ |

Table 1: Summary of the values of the physical parameters used for various radial-azimuthal test-case simulations

All test-case simulations are run for 30 $\mu$s, expect for Case 1 (the baseline) for which the simulation time is 135 $\mu$s. The set of values of the physical parameters $S$, $E$, and $B$ for Case 1 correspond to the conditions of the radial-azimuthal benchmark [23].

## 2.2. Axial-azimuthal Hall-thruster-representative configuration

The simulations' setup is similar to the one presented in Refs. [17][18][22]. The domain is a 2D Cartesian ($x - z$) plane with the dimensions of 2.5 × 1.28 cm. The $x$-axis is along the axial coordinate and the $z$-axis is along the azimuth. Other details of the setup such as the cell size, time step, total simulation time, and the initial plasma and boundary conditions are the same as those reported in Refs. [18][22].

The axial-azimuthal simulations feature an imposed temporally invariant cosine-shaped ionization source with the peak value corresponding to the total axial ion current density ($J$), which is set as a simulation parameter. Our various axial-azimuthal test cases, thus, differ from each other based on the value of $J$. Table 2 provides the values of total ion current density for each of the test cases.

| Test case | J value |
|---|---|
| 1 | $1/8\ J_0 = 50\ Am^{-2}$ |
| 2 | $1/4\ J_0 = 100\ Am^{-2}$ |
| 3 | $1/2\ J_0 = 200\ Am^{-2}$ |
| 4 (Baseline) | $J_0 = 400\ Am^{-2}$ |

Table 2: Summary of values of $J$ corresponding to various radial-azimuthal test-case simulations; the baseline case has the $J$ value equal to that of the axial-azimuthal benchmark [22].

The ground-truth data for the axial-azimuthal test cases are obtained from reduced-order simulations with a domain decomposition of $M = 40$ and $N = 20$. Compared to full-2D simulations, the quasi-2D ones have, thus, a computational gain by about a factor of 6 as was demonstrated in Ref. [17].

## Section 3: Dynamics prediction results and discussion

In this section, we present and discuss the predictions of the linear reduced-order models derived from the application of the OPT-DMD to the data from the radial-azimuthal and axial-azimuthal test cases. It is important to note that the results shown in the following correspond to the cases with the most insightful and distinct outcomes among all test cases on which the OPT-DMD algorithm was trained and tested.

Throughout this section, the performance of the OPT-DMD reduced-order models is first assessed through comparing the ground-truth and the predicted temporal evolutions of the spatially averaged plasma properties in each test case. Second, the 2D snapshots of plasma properties from the OPT-DMD model for each test case are juxtaposed against the snapshots from the corresponding ground-truth simulation at sample time instants within the test window. This provides a visual means of comparing the extent of similarity between the OPT-DMD models' prediction and the ground-truth data.

The difference between each ground-truth snapshot from the PIC simulation and the predicted one from the OPT-DMD model is also quantified in terms of a loss factor ($\mathcal{L}$) defined as per Eq. 1.

$$\mathcal{L} = \frac{\|S^{true} - S^{DMD}\|_F}{\|S^{true}\|_F}, \quad \text{(Eq. 1)}$$

In Eq. 1, $\|S\|_F$ denotes the Frobenius norm of the data matrix that represents a 2D plasma property snapshot from either the ground-truth ($\|S^{true}\|_F$) or the OPT-DMD model ($\|S^{DMD}\|_F$). The Frobenius norm is calculated as in Eq. 2,

$$\|S\|_F \equiv \sqrt{\sum_{i=1}^{m} \sum_{j=1}^{n} |s_{ij}|^2}, \quad \text{(Eq. 2)}$$

where, $m$ and $n$ are the number of grid points along the $x$ and $z$ directions, respectively, and $s_{ij}$ represents the value of the plasma property at each node $ij$.

### 3.1. Radial-azimuthal test case

We start off by presenting a number of radial-azimuthal test cases in which the OPT-DMD models demonstrate remarkably good predictiveness. In this regard, Figure 1 shows the time evolutions of the spatially averaged plasma properties from the ground-truth PIC simulation (black line) and the OPT-DMD model (red line) for the test cases with $E = 0.5E_0$ and $E = 5E_0$.

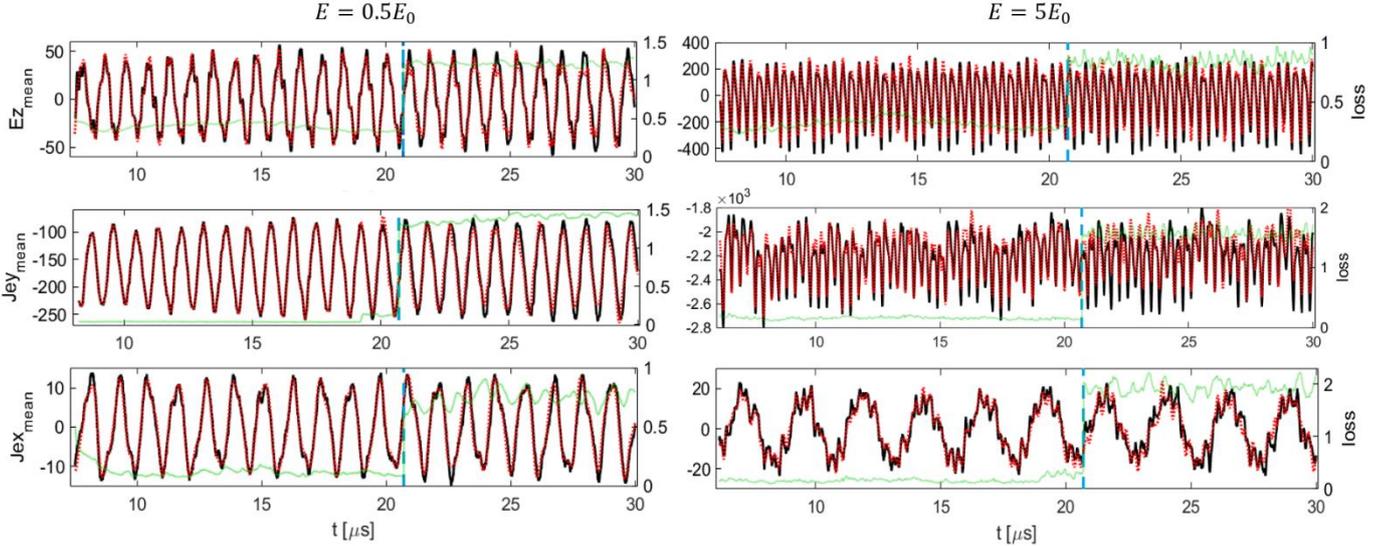

Figure 1: Time evolution of the spatially averaged plasma properties for two test cases with the axial electric field values of $E = 0.5E_0$ (**left column**), and $E = 5E_0$ (**right column**). First-row plots represent the azimuthal electric field ($E_z$) signal, the second-row plots show the axial electron current ($J_{ey}$), and the third-row plots correspond to the radial electron current ($J_{ex}$). Ground-truth traces from the radial-azimuthal quasi-2D simulations are shown with solid black lines, OPT-DMD predicted traces with dotted red lines, and the green traces represent the time variation of the loss factor (Eq. 1). Dashed blue lines indicate the training end time.

In each case, and for all plasma properties, we have used the data in the time window of about 5-22 $\mu s$ for the training of the OPT-DMD model, which was subsequently tested over the timeframe of about 22-30 $\mu s$. Across the two cases and all plasma properties, we have used a truncation rank ($r$) of 50 to 60 in the OPT-DMD. The data within the first 5 $\mu s$ of the discharge evolution were discarded for the training as it represents the initial transient of the system.

We observe that, in terms of the time evolution of the plasma properties, the traces corresponding to the OPT-DMD models in Figure 1 closely follow the ground-truth traces within both the training and testing time windows. Moreover, the time evolution of the loss factor ($\mathcal{L}$), illustrated as green lines in the plots of Figure 1, shows an expected sudden variation when moving from the training window to the test but it remains stable and bounded for both the training and the test datasets.

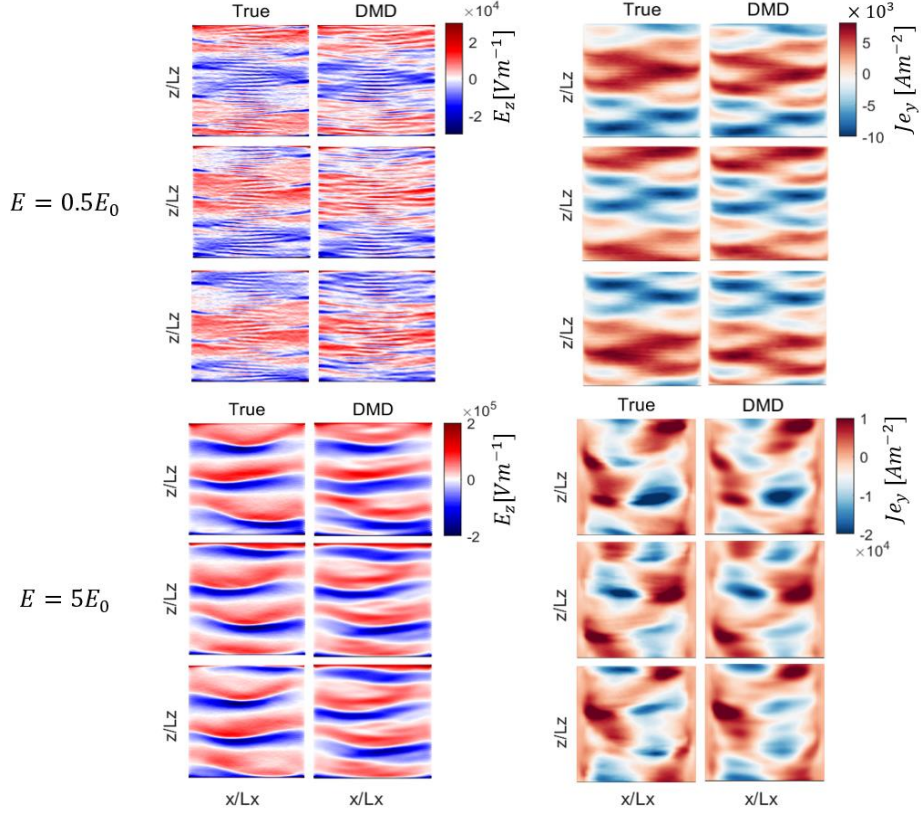

Figure 2: Comparison of the 2D snapshots of the plasma properties from the ground-truth radial-azimuthal quasi-2D PIC simulations and the OPT-DMD model for two values of the axial electric field at three randomly selected time instants within the test window ($\sim 22 - 30$ $\mu s$). **Left column**: azimuthal electric field ($E_z$), **right column**: axial electron current ($J_{ey}$). The first three rows correspond to $E = 0.5E_0$, whereas the second three rows are for the simulation with $E = 5E_0$.

In Figure 2, we have compared the 2D snapshots of the azimuthal electric field ($E_z$) and the axial electron current density ($J_{ey}$) from the ground-truth simulations and the OPT-DMD models for the test cases of $0.5E_0$ and $5E_0$ at three randomly selected time instants within the test window. The predicted snapshots of either plasma property are seen to be closely representative of the ground-truth ones in both test cases and at all three time instants. In any case, the $J_{ey}$ snapshots present a higher degree of similarity between the predicted and the ground-truth data in terms of both the spatial structure and the magnitudes.

Another radial-azimuthal test case in which the derived model from the OPT-DMD exhibits great predictive capability is the case with $B = 1.5B_0$. The comparison of the OPT-DMD model predictions against the ground-truth for this test case in terms of the time evolution of the averaged plasma properties and their 2D snapshots is shown in Figure 3. Similar to the test cases with $E = 0.5E_0$ and $5E_0$, we have used the time window of about 5-22 $\mu s$ for the training and an $r$ value of 50-60 across different plasma properties.

It is observed from Figure 3 that the OPT-DMD prediction within the test window are highly consistent with the ground-truth data with regard to the time evolutions of the plasma properties as well as the 2D snapshots at sample time instants. It is also noteworthy that, in this test case, the variation in $\mathcal{L}$ from the training window to the test is more gradual and does not feature the step change that was observed for the $0.5E_0$ and $5E_0$ test cases in Figure 1. In any case, the loss factor variation over time is again seen to be bounded.

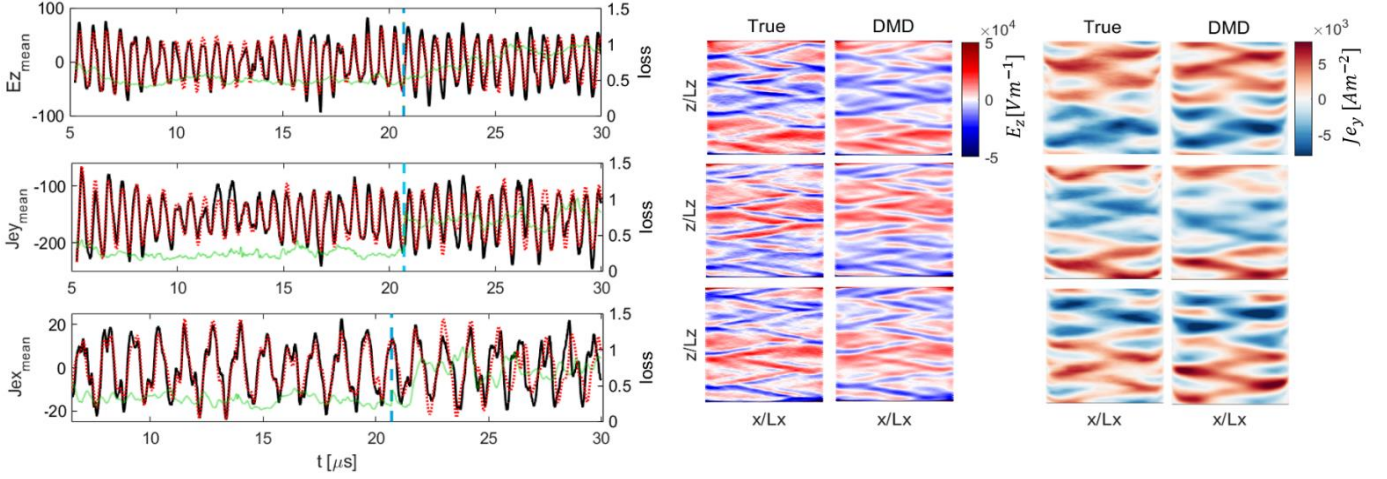

Figure 3: (**Left column**) time evolution of the spatially averaged plasma properties, $E_z$, $J_{ey}$, and $J_{ex}$, for a test case with the radial magnetic field intensity of $B = 1.5B_0$. Ground-truth traces from the quasi-2D simulation are shown with solid black lines, OPT-DMD predicted traces with dotted red lines, and the green traces represent the time variation of the loss factor (Eq. 1). Dashed blue lines indicate the training end time. (**Middle column**) Comparison of the 2D snapshots of $E_z$ from the ground-truth simulation and the OPT-DMD model at three randomly selected time instants within the test window ($\sim 22 - 30 \ \mu s$). (**Right column**) Comparison of the 2D snapshots of $J_{ey}$ from the ground-truth simulation and the OPT-DMD model at three randomly selected time instants within the test window.

For the test cases presented so far, a common feature in the time evolution plots of the averaged plasma properties (Figure 1 and Figure 3(left column)) was that the signals were rather periodic. This implies that, under the setup conditions corresponding to these test cases, the system exhibits a relatively quasi-steady behavior after its initial transient.

Nonetheless, we will demonstrate in the following in several test cases that the predictive performance of the linear OPT-DMD models is to some extent reduced for the cases where the signal evolution becomes less regular, i.e., the plasma system is characterized by underlying transient processes and highly nonlinear interactions.

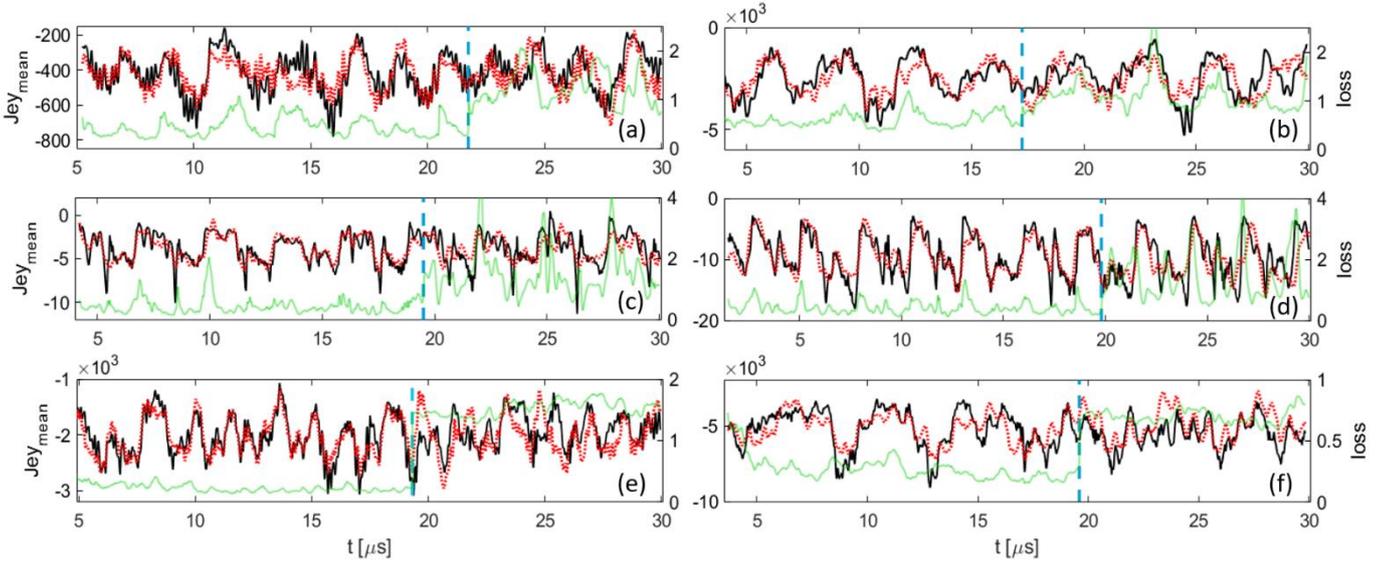

Figure 4: Spatially averaged axial electron current density ($J_{ey}$) signal for radial-azimuthal test cases with various setup parameters: (a) baseline conditions with axial electric field ($E$) being $E_0$, and peak amplitude of the ionization source ($S$) being $S_0$. (b) $E = 3E_0$, $S = S_0$, (c) $E = E_0$, $S = \frac{S_0}{16}$, (d) $E = E_0$, $S = \frac{S_0}{8}$, (e) $E = E_0$, $S = 3S_0$, and (f) $E = E_0$, $S = 6S_0$. Ground-truth traces from the quasi-2D simulations are shown with solid black lines, OPT-DMD predicted traces with dotted red lines, and the green traces represent the time variation of the loss factor (Eq. 1). Dashed blue lines indicate the training end time.

In this respect, Figure 4 compares the time evolution signals of the axial electron current density from the PIC simulations and the OPT-DMD models for six different sets of setup parameters that correspond to the test cases 1, 3, and 6-9 in Table 1. Noting the specificities of the discharge evolution in each test case, which is observable as different temporal variation characteristics of the spatially averaged $J_{ey}$ across the test cases, the end time of the training window varies between about 17 to 22 $\mu$s for various cases and the $r$ value used has been between 100-150.

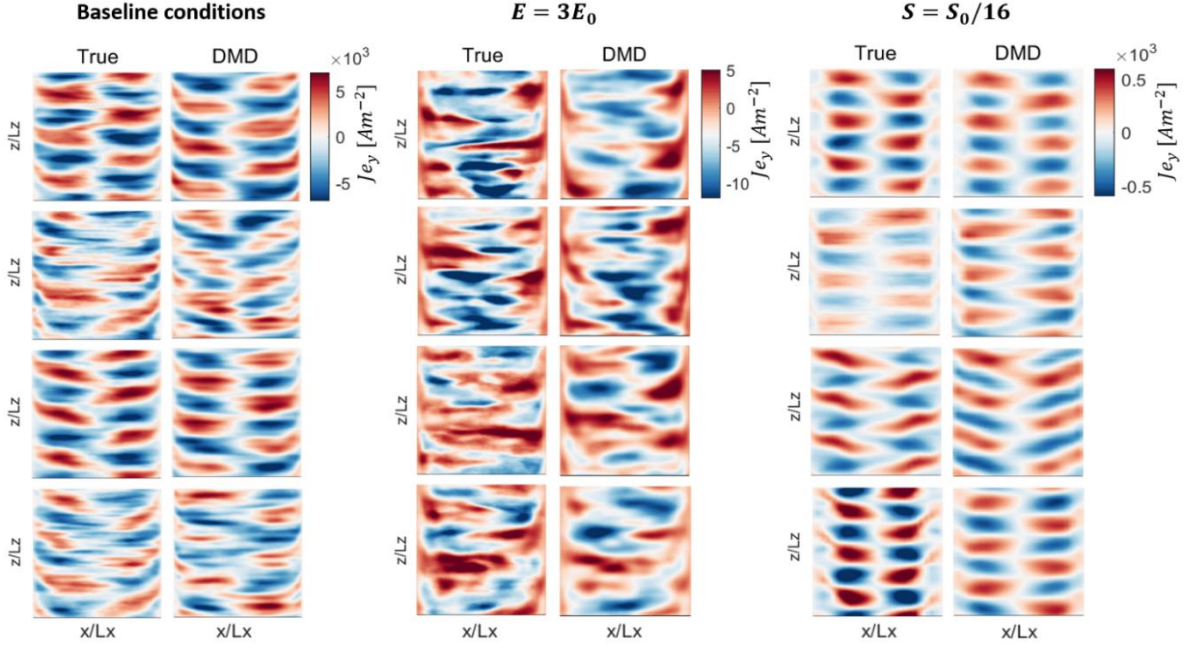

Figure 5: Comparison between the 2D snapshots of the axial electron current density ($J_{ey}$) from the ground-truth quasi-2D PIC simulations and the OPT-DMD model for three sets of setup parameters; (**left column**) baseline conditions, $E = E_0$, $S = S_0$, (**middle column**) $E = 3E_0$, $S = S_0$, (**right column**) $E = E_0$, $S = \frac{S_0}{16}$. For each set of setup parameters, the snapshots along the rows correspond to various instances of the system evolution within the associated test window for those setup parameters (from the dashed blue line in Figure 4 until t = 30 $\mu$s).

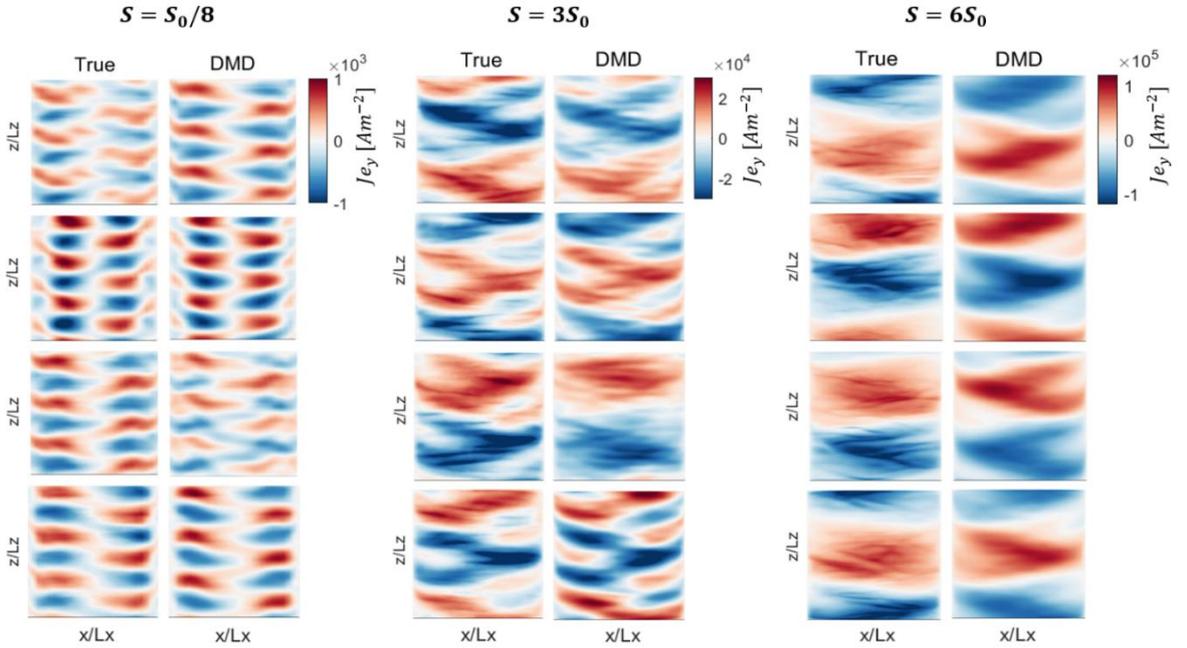

Figure 6: Comparison between the 2D snapshots of the axial electron current density ($J_{ey}$) from the ground-truth quasi-2D PIC simulations and the OPT-DMD model for three sets of setup parameters; (**left column**) $E = E_0$, $S = \frac{S_0}{8}$, (**middle column**) $E = E_0$, $S = 3S_0$, (**right column**) $E = E_0$, $S = 6S_0$. For each set of setup parameters, the snapshots along the rows correspond to various instances of the system evolution within the associated test window for those setup parameters (from the dashed blue line in Figure 4 until t = 30 $\mu$s).

We can observe in Figure 4 that, even though the OPT-DMD predictions may not be as perfect as those shown for test cases 2, 4, and 5 (Table 1), the simple, linear, and highly low computational cost models obtained from the OPT-DMD algorithm still perform a decent job in recovering and predicting the time evolution of the $J_{ey}$ within the training and testing windows, respectively. In addition, the loss factor is noticed to still remain bounded for both the training and the testing.

Furthermore, referring to Figure 5 and Figure 6, it is additionally seen that, despite some differences observed between the time evolution signals of $J_{ey}$ from the OPT-DMD and the ground-truth simulations, the predicted 2D snapshots at four random time instants within the test window show a high degree of resemblance to the corresponding ground-truth data for all of test cases 1, 3, and 6 to 9. Indeed, we observe that the main features in the ground-truth data are captured by the OPT-DMD models, and the magnitudes are very similar as well.

### 3.1.1. An example of long-term dynamics prediction

At this point, one may wonder about the stability and the predictive capability of an OPT-DMD model for long-term predictions of the dynamics. We have assessed this for test case 1 (Table 1), the setup conditions for which represent those of the radial-azimuthal benchmark [23].

We have run the quasi-2D PIC simulation of test case 1 for 135 $\mu$s and have evaluated the performance of three OPT-DMD models obtained with different values of the truncation rank (or, equivalently, the number of DMD modes), namely, $r = 55$, $r = 150$, and $r = 400$. For the cases with $r = 55$ and $r = 150$, the training time window is from about 5 to 22 $\mu$s. For the case with $r = 400$, we needed to extend the end time of the training window to about 97 $\mu$s. This was necessary because the relatively large truncation rank of 400 requires sufficient number of training data snapshots as well so that the corresponding model will still be learning the physical patterns in the data (rather than the noise) and will thus not overfit the training data.

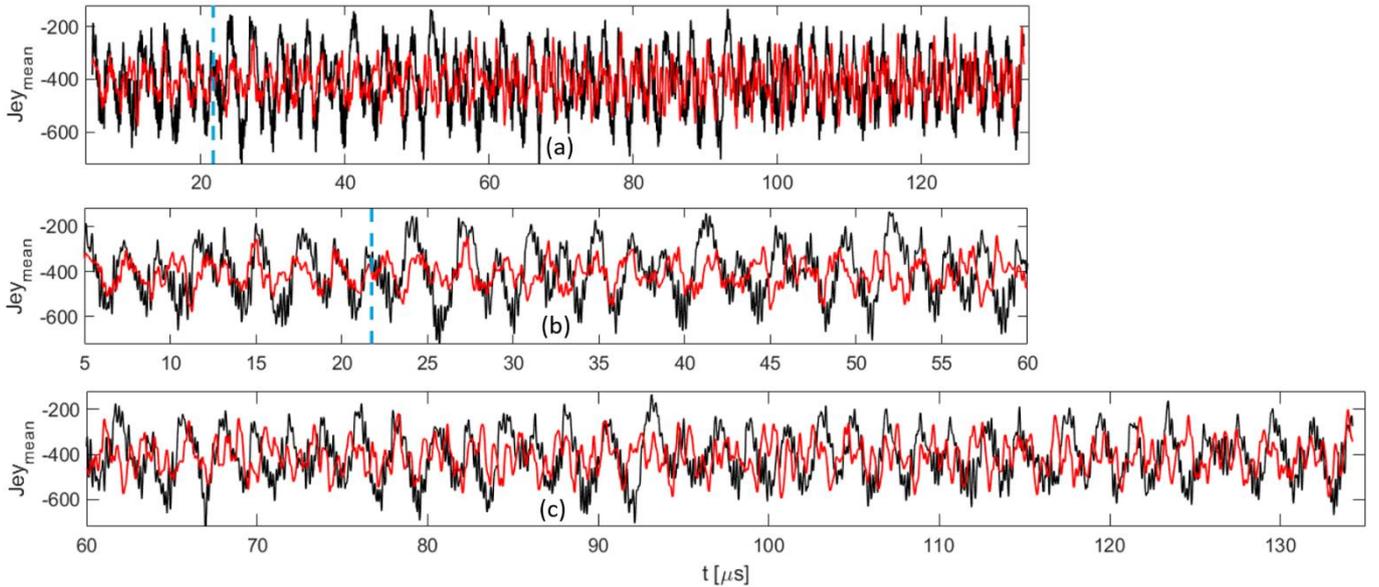

Figure 7: (a) Time evolution of the spatially averaged axial electron current density ($J_{ey}$) from the quasi-2D PIC simulation (solid black line) and the OPT-DMD model (solid red line) with the rank of $r = 55$. (b) zoomed view on plot (a) in the time window of 5-60 $\mu s$. (c) zoomed view on plot (a) in the time window of 60-135 $\mu s$. The setup parameters correspond to those of the baseline radial-azimuthal test case. Dashed blue lines indicate the training end time.

Figure 7 shows the evolution of the spatially averaged $J_{ey}$ from the ground-truth PIC simulation and the OPT-DMD model with $r = 55$ over the timeframe of 5-135 $\mu$s. It is observed that, even with a training-to-test ratio of about 0.15, the predicted signal from the OPT-DMD model remains fully stable in time and it follows the general trends of the ground-truth throughout the test window. The stability of the predicted signal from OPT-DMD is indeed remarkable in itself because, as discussed in detail in Part I of the article, obtaining a stable and, hence, reliable DMD model is quite often challenging with the original version of the DMD method (basic/exact DMD).

When increasing the number of DMD modes from 55 to 150, the time evolution plots of averaged $J_{ey}$ in Figure 8 illustrate that, whereas the model now better fits the training, its prediction within the test window follows less closely the ground-truth compared to the model with $r = 55$. The predicted signal still has the same mean value as that of the ground-truth and remains stable over time, but it is overall out of sync with the signal from the PIC simulation.

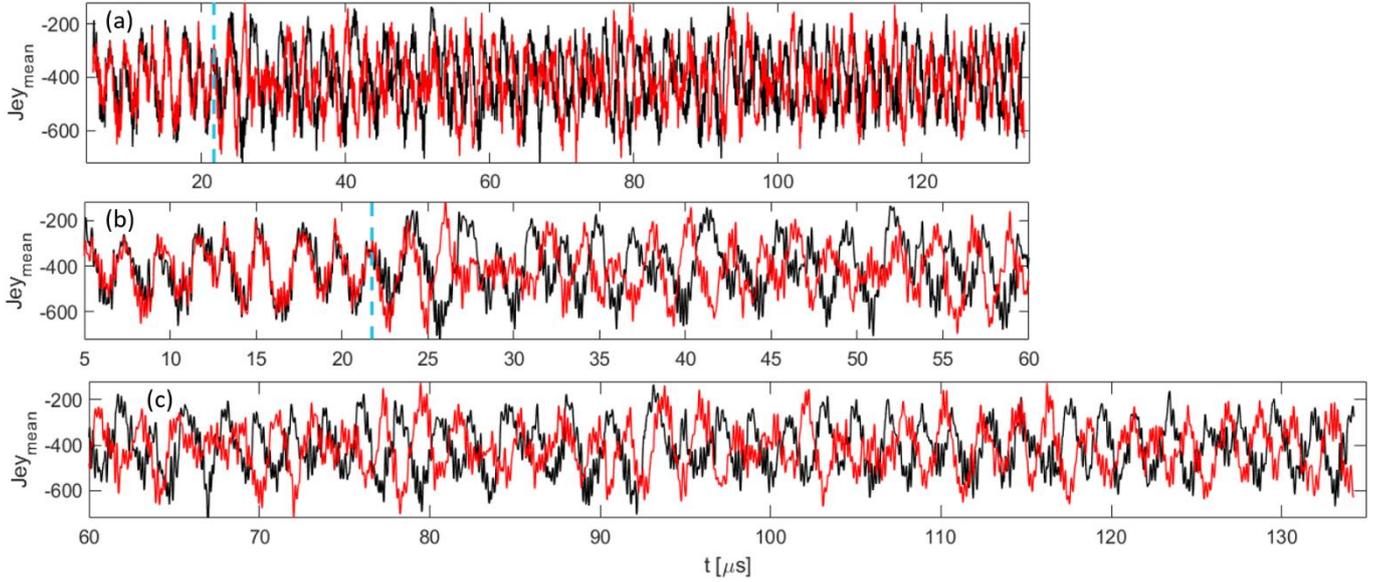

Figure 8: (a) Time evolution of the spatially averaged axial electron current density ($J_{ey}$) from the quasi-2D PIC simulation (solid black line) and the OPT-DMD model (solid red line) with the rank of $r = 150$. (b) zoomed view on plot (a) in the time window of 5-60 $\mu s$. (c) zoomed view on plot (a) in the time window of 60-135 $\mu s$. The setup parameters correspond to those of the baseline radial-azimuthal test case. Dashed blue lines indicate the training end time.

In the case of $r = 400$, the averaged $J_{ey}$ signal from the OPT-DMD model is noticed from Figure 9 to capture the main evolution trend of the training data, and this behavior is also exhibited by the predicted signal into the test window. However, we can also observe that the prediction falls frequently out-of-sync with the ground-truth signal.

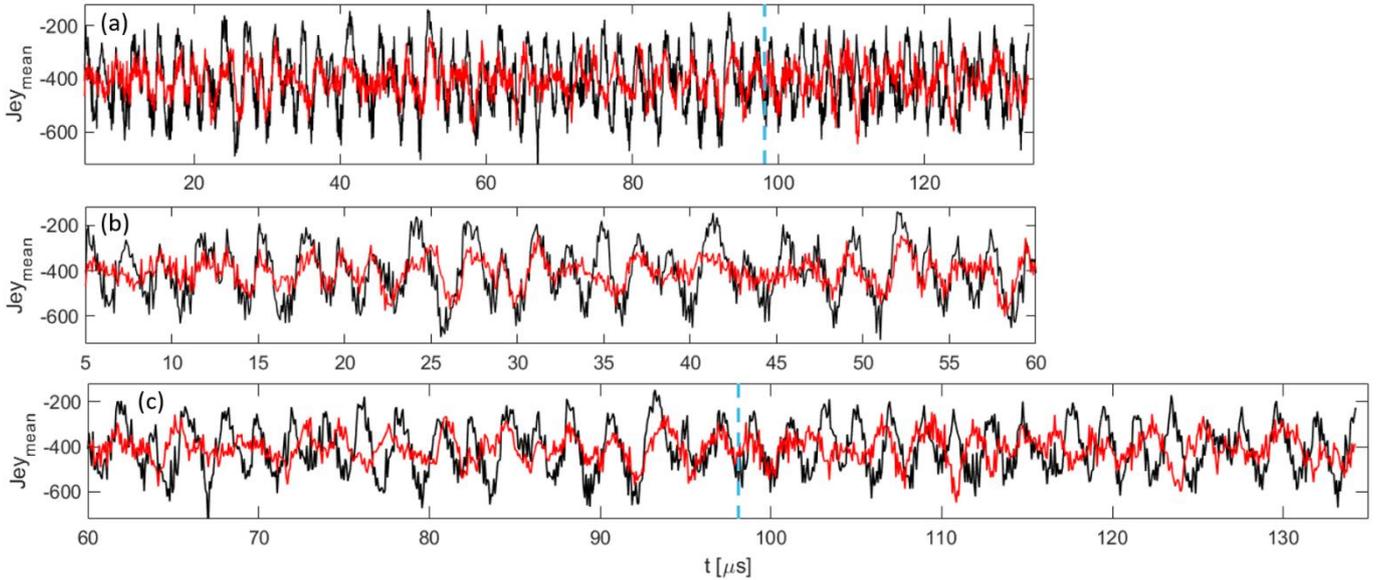

Figure 9: (a) Time evolution of the spatially averaged axial electron current density ($J_{ey}$) from the quasi-2D PIC simulation (solid black line) and the OPT-DMD model (solid red line) with the rank of $r = 150$. (b) zoomed view on plot (a) in the time window of 5-60 $\mu s$. (c) zoomed view on plot (a) in the time window of 60-135 $\mu s$. The setup parameters correspond to those of the baseline radial-azimuthal test case. Dashed blue lines indicate the training end time.

The time evolutions of the loss factor ($\mathcal{L}$) in $J_{ey}$ between the ground-truth data and the results of the OPT-DMD models with different $r$ values are shown in Figure 10. It is again observed that, in all cases, the loss remains bounded, which is particularly interesting for the models with $r = 55$ and $150$ that were tested over a notably extended timeframe.

We would point out that comparing the values of $\mathcal{L}$ across different models might be rather misleading. In fact, it is recalled that increasing the truncation rank of the DMD enables the model to capture more of the patterns underlying the training data. This reduces the loss factor, which is calculated by comparing the ground-truth and reconstructed snapshots point-by-point. However, as seen from the $J_{ey}$ time evolution plots in Figure 7 and Figure 8, a higher rank model does not necessarily perform better in terms of the time variation of the spatially averaged data in the testing window. This can be in part explained by the fact that the DMD assumes the underlying modes to remain invariant over time. In this regard, the higher the number of DMD modes retained, i.e., for larger $r$ values, the more probable it is in real nonlinear systems that feature transient behaviors that some of the learned modes would not exist within the training set. This can consequently impact the ability of the model to properly trace the evolution trends in the ground-truth.

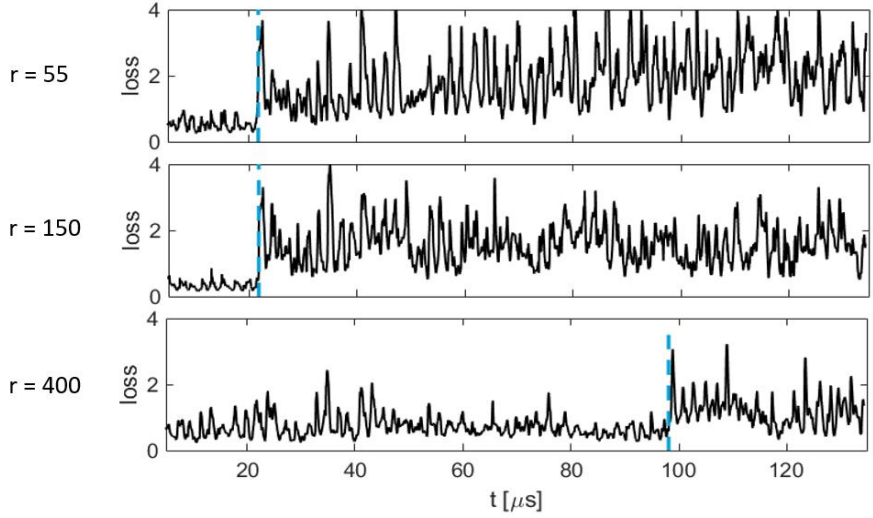

Figure 10: Evolution of the loss factor (Eq. 1) in the axial electron current density ($J_{ey}$) signal between the OPT-DMD model with different values the truncation rank ($r$) and the quasi-2D PIC simulation of the baseline radial-azimuthal test case. Dashed blue lines indicate the training end time.

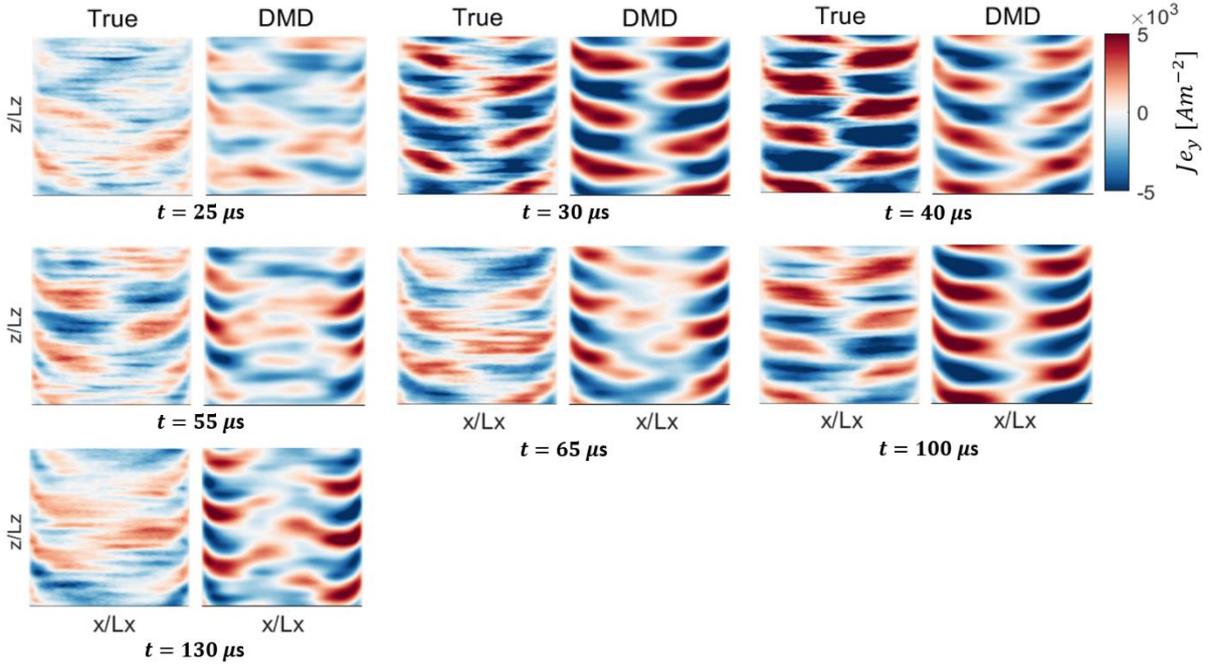

Figure 11: Comparison over several test-window time instants between the 2D snapshots of the axial electron current density ($J_{ey}$) from the ground-truth quasi-2D simulation and the OPT-DMD model with the rank of $r = 55$. The setup parameters correspond to those of the baseline radial-azimuthal test case.

To conclude this subsection, we refer to Figure 11 and Figure 12 in which, for the OPT-DMD models with $r = 55$ and 400, respectively, we have compared the predicted 2D snapshots of the $J_{ey}$ against the ground-truth at several test time instants. It is noted that, generally in both cases, the predicted snapshots feature the main patterns, and the magnitudes are also overall comparable. However, the OPT-DMD model with $r = 400$ has expectedly captured the finer details of the spatial distribution, which in turn has resulted in an increased degree of resemblance between the DMD-predicted and the ground-truth data in this case.

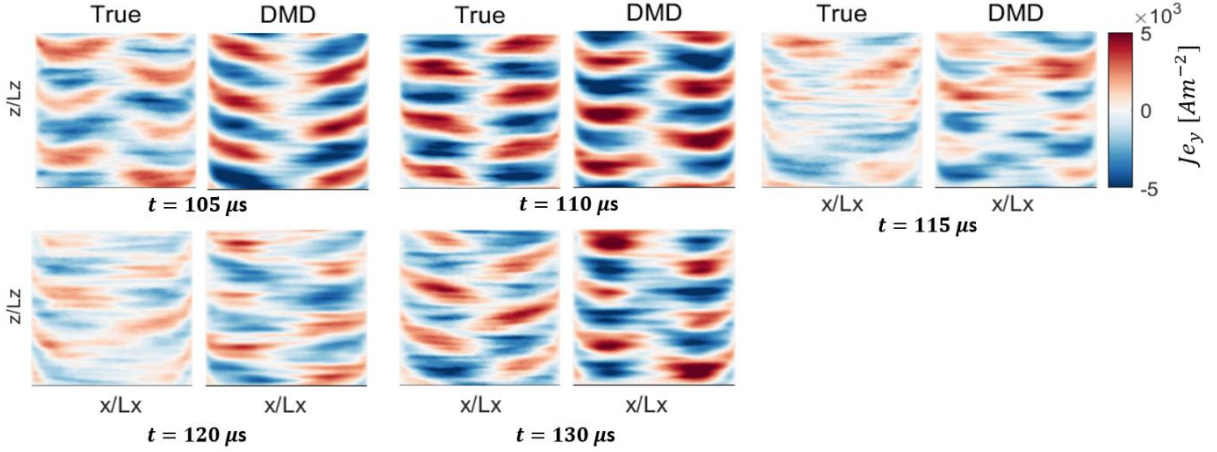

Figure 12: Comparison over several test-window time instants between the 2D snapshots of the axial electron current density ($J_{ey}$) from the ground-truth quasi-2D simulation and the OPT-DMD model with the rank of $r = 400$. The setup parameters correspond to those of the baseline radial-azimuthal test case.

### 3.2. Axial-azimuthal configuration

To present the performance of the OPT-DMD models in the axial-azimuthal test cases, we first look at the time evolution plots of the spatially averaged azimuthal electric field ($E_z$) shown in Figure 13 for various values of the total ion current density ($J$). We compare in this figure the ground-truth and the OPT-DMD-predicted $E_z$ signal for different test cases. We have also superimposed on the plots the time variation in the loss factor between the ground-truth and the reconstructed data.

For all test cases shown in Figure 13, the DMD truncation rank of 55 has been used. Moreover, for test cases 1 to 3 (Table 2) corresponding to the $J$ values of 50 to 200 $Am^{-2}$, the training time window is about 18 $\mu s$. For test case 4 ($J = 400\ Am^{-2}$), we have instead used a slightly shorter training time window of about 16 $\mu s$.

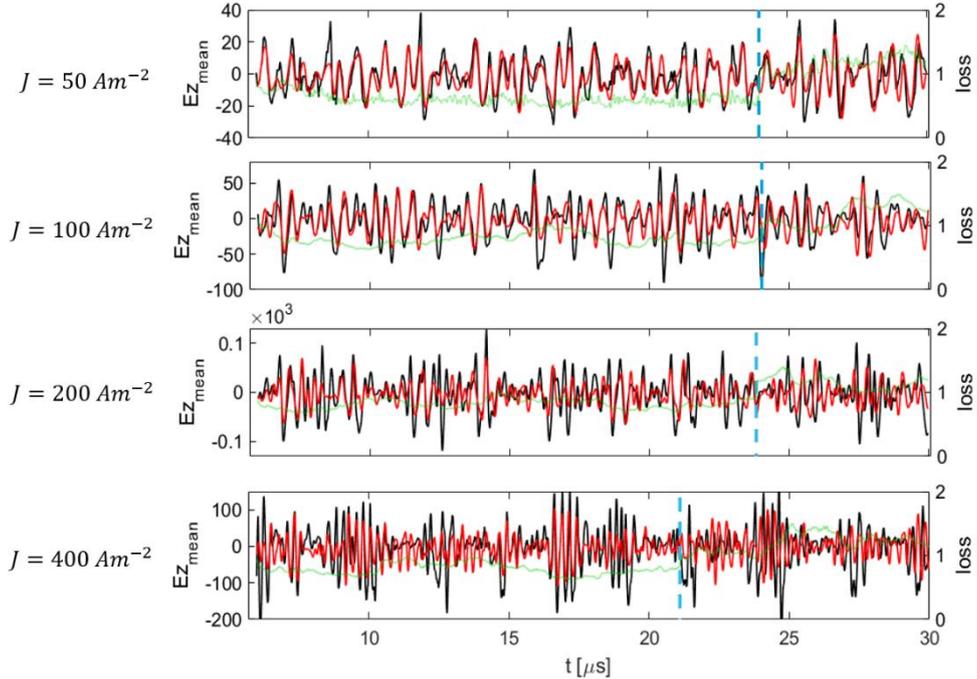

Figure 13: Time evolution of the spatially averaged azimuthal electric field ($E_z$) for the axial-azimuthal test cases with four different values of the total ion current density ($J$). Ground-truth traces from the quasi-2D simulations are illustrated as solid black lines and the OPT-DMD predicted traces as solid red lines. The green traces represent the time variation of the loss factor (Eq. 1). Dashed blue lines indicate the training end time.

It is evident from the time evolution plots in Figure 13 that the $E_z$ signal becomes increasingly more complex with the increasing total ion current density. As it has been seen in our previous works [18][24] as well as in the literature [25][26], the higher complexity of the $E_z$ signal at higher $J$ values roots in the presence of underlying nonlinear and transient processes at these values of $J$, such as the development of the Electron Drift Instability (EDI) and its transition to the ion acoustic waves. Nevertheless, the linear models from the OPT-DMD are observed to provide reasonable predictions of the ground-truth data evolution, especially at lower ion current densities of 50 and 100 $Am^{-2}$, where the OPT-DMD-predicted traces follow the ground-truth very closely. The loss factor also varies quite gradually in time from the training to the test window, and it remains stable and bounded throughout the training and testing as was the case for the radial-azimuthal test cases.

For the axial-azimuthal test case 1 ($J = 50\ Am^{-2}$), where the OPT-DMD model was noticed to present a great predictive capability at DMD rank of 55, we have shown in Figure 14 the effect of the length of the training time window by comparing the time evolution and the 2D snapshots of the azimuthal electric field from the ground-truth PIC simulation and the OPT-DMD model.

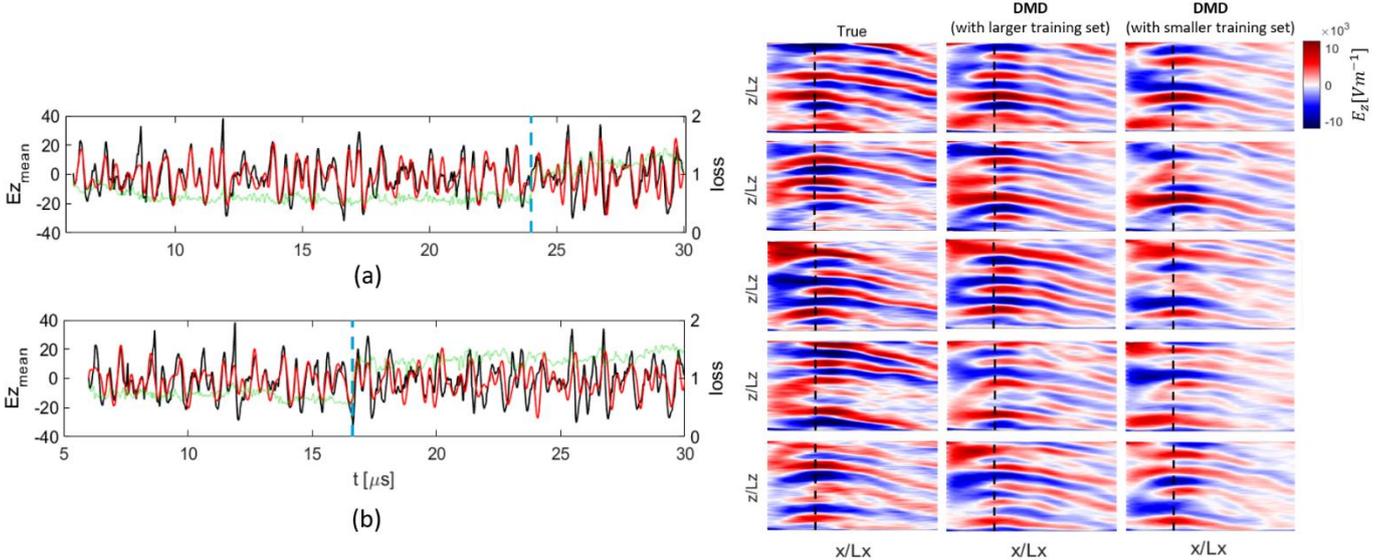

Figure 14: (**Left column**) time evolution of the spatially averaged azimuthal electric field ($E_z$) for the axial-azimuthal test case with $J = 50\ Am^{-2}$ and two different lengths of the training window: (a) larger training set of $6 - 24\ \mu s$, (b) smaller training set of $6 - 17\ \mu s$. Ground-truth traces from the quasi-2D simulation are illustrated as solid black lines and the OPT-DMD predicted traces as solid red lines. The green traces represent the time variation of the loss factor (Eq. 1). Dashed blue lines indicate the training end time. (**Right column**) Comparison between the 2D snapshots of $E_z$ from the ground-truth PIC simulation and the OPT-DMD model obtained with the larger and smaller training windows.

Referring to the time evolution plots (Figure 14(a) and (b)), we observe that, with a larger training set, the OPT-DMD model can more closely predict the ground-truth signal. This is further evident by referring to the comparison of the snapshots at four random test time instants. It is indeed clear that the predicted 2D snapshots from the model trained on a larger dataset are quite more similar to the ground-truth snapshots, particularly in terms of the details of the spatial distribution of the azimuthal electric field.

In any case, the model trained on a smaller dataset also captures the main spatial patterns in the data, and its predicted $E_z$ signal largely follows the overall trend of the ground-truth data evolution as well.

**Section 4: Conclusions**

In this paper, we assessed the applicability of the OPT-DMD method to provide reliable and predictive reduced-order models of plasma systems. In this regard, we evaluated the performance of the linear data-driven models from the OPT-DMD to predict the temporal evolution of the plasma state and the spatial distribution of the properties in various test cases representative of radial-azimuthal and axial-azimuthal cross-sections of a Hall thruster. We overall observed that the OPT-DMD ROMs have a promising predictive potential across all test cases, with a remarkable ability to predict the ground-truth data in cases where the plasma system exhibits relatively quasi-periodic behavior. We additionally noted that the OPT-DMD models provide stable predictions over time, and that the loss factor between the ground-truth and the reconstructed data remains bounded throughout the training and the test windows in all cases tested.

In the majority of test cases, either in the radial-azimuthal or the axial-azimuthal configuration, we examined the short-time predictive capacity of the OPT-DMD models because this is the most prudent use of a machine-learning algorithm such as Dynamic Mode Decomposition that provides linear dynamics models under the assumption that the underlying patterns (modes) in the training data remain invariant over time. This can be a stringent assumption for real-world plasma systems such as Hall thrusters and the broader family of E×B discharges where nonlinear, highly transient processes and interactions often take place.

In any case, for a radial-azimuthal test case, we investigated the behavior of the OPT-DMD model for long-term prediction of the dynamics. We noticed that models developed using three different truncation ranks of the DMD all present great stability and can acceptably reproduce the time evolution trends in the ground-truth data as well as the spatial structures in the 2D distributions of the plasma properties.

Another noteworthy aspect is that, in this article, we mostly focused on the qualitative assessment of the OPT-DMD models, noting the scope and the context of the research at this stage. We did not report detailed analyses on the influence of hyper-parameters such as the truncation rank ($r$) on the derived OPT-DMD models. These investigations are left for future work where, building upon the results of this first-of-its-kind effort, we can devise strategies and/or algorithms to enable the fine-tuning of the hyper-parameters.

As the final remark, it is important to put the results reported in this article on the OPT-DMD application to ML-enabled plasma modelling in an applied context. Indeed, when developing reduced-order models using machine learning, the model use cases shall be kept in mind as the computational cost and the accuracy of the model often come as a trade-off. The OPT-DMD models are of negligible computational cost when compared to the PIC simulations, enabling very fast predictions of the plasma behavior. However, their degree of accuracy can be different across various plasma configurations and conditions depending on the complexity of interactions between the underlying physical processes. In any case, our results showed that, in general, the OPT-DMD can be a reliable method to derive models for applications such as operational control and/or digital system model/twin development where the models are intended to be predictive over short timeframes corresponding to the acquisition and receipt of new datasets from the physical system that can enable updating the ML model.

As we progress toward the overarching goal of establishing digital technologies for cross-field plasma systems, the results of this two-part article indicate that the OPT-DMD method can serve as one of the underpinning numerical capabilities.

**Acknowledgments**:

The present research is carried out within the framework of the project "Advanced Space Propulsion for Innovative Realization of space Exploration (ASPIRE)". ASPIRE has received funding from the European Union's Horizon 2020 Research and Innovation Programme under the Grant Agreement No. 101004366. The views expressed herein can in no way be taken as to reflect an official opinion of the Commission of the European Union.

MR, FF, and AK gratefully acknowledge the computational resources and support provided by the Imperial College Research Computing Service (http://doi.org/10.14469/hpc/2232).

**Data Availability Statement**:

The simulation data that support the findings of this study are available from the corresponding author upon reasonable request.